\documentclass{PoS}

\usepackage{graphicx}
\usepackage{amsmath,amssymb,amsfonts}
\usepackage{subfigure}
\usepackage{listings}
\usepackage{cite}

\newcommand{\secdec}{{\textsc{SecDec}}}
\def\be{\begin{equation}}
\def\ee{\end{equation}}
\newcommand{\bea}{\begin{eqnarray}}
\newcommand{\eea}{\end{eqnarray}\noindent}

\newcommand{\bcen}{\begin{center}}
\newcommand{\ecen}{\end{center}}

\def\url#1{\texttt{#1}}
\def\eps{\epsilon}

\title{Two-loop applications of the program SecDec}

\ShortTitle{Applications of SecDec}

\author{\speaker{Sophia Borowka}\\
       Max Planck Institute for Physics, F\"ohringer Ring 6, 80805 Munich, Germany\\
       E-mail: \email{sborowka@mpp.mpg.de}}

\author{Gudrun Heinrich\\
        Max Planck Institute for Physics, F\"ohringer Ring 6, 80805 Munich, Germany\\
        E-mail: \email{gudrun@mpp.mpg.de}}

\abstract{
\secdec{} is a program which can be used for the 
factorisation of poles and subsequent 
numerical evaluation of multi-loop integrals, in particular massive two-loop integrals.
We show applications to two-loop master integrals entering the calculation of 
top quark pair production at NNLO, and to the dominant momentum dependent two-loop corrections 
to the neutral Higgs boson masses in the MSSM.
}

\FullConference{11th International Symposium on Radiative Corrections (Applications of Quantum Field Theory to Phenomenology) (RADCOR 2013),\\
		22-27 September 2013\\
		Lumley Castle Hotel, Durham, UK }

\begin{document}
\section{Introduction}

Precision measurements are of primary importance in order to 
find deviations from the Standard Model and to identify their origin as 
``New Physics".
Certainly, this is only possible in conjunction with  precise theory predictions,  
requiring the calculation of higher order corrections.
While the calculation of corrections at next-to-leading order 
has reached an impressive level of automation meanwhile, 
corrections beyond one loop still require quite some efforts 
both on the conceptual and on the technical side before 
they can  be produced in a largely automated way.

An important ingredient for the calculation of higher order corrections are
the loop integrals entering the virtual corrections.
Analytical expressions for integrals beyond one loop are only known 
for integrals depending on rather few mass scales.
Therefore, as soon as several mass scales are involved, numerical methods 
to calculate these integrals are often indispensable.
However, the latter can only be applied after extraction of possible 
ultraviolet and/or infrared poles contained in the diagrams.

The program \secdec{}\,\cite{Carter:2010hi,Borowka:2012yc,Borowka:2013cma}, performs the task of 
isolating dimensionally regulated singularities in an automated way, 
based on the algorithm of 
sector decomposition \cite{Binoth:2000ps,Roth:1996pd,Hepp:1966eg}.
Other implementations of sector decomposition into public programs are also available
\cite{Smirnov:2009pb,Bogner:2007cr,Gluza:2010rn}. 
However, the latter are more or less restricted to the Euclidean region, 
while \secdec{} can deal with physical kinematics, including for instance mass thresholds.

In these proceedings, applications of  the 
program \secdec\,2.1 to massive two-loop integrals will be presented.


\section{Structure of the program}
\label{sec:structure}

\begin{figure}[htb]
\includegraphics[width=14cm]{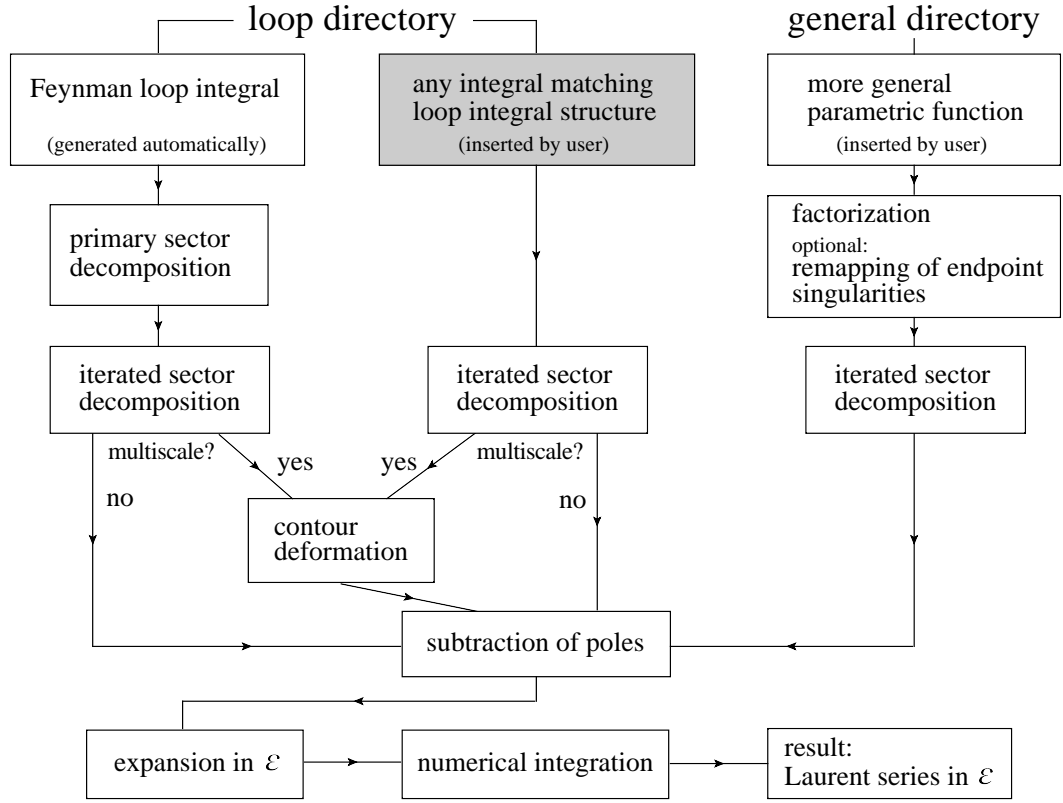}
\caption{Flowchart showing the structure of the program \secdec{}.}
\label{fig:flowchart}
\end{figure}

The program can be divided into two 
main branches, one for the computation of loop integrals, the other one for the 
treatment of more general parametric functions
(corresponding to the directories {\it loop} and {\it general}).
A flowchart of the program is depicted in Fig.~\ref{fig:flowchart}. 
For a detailed description of the program we refer to \cite{Carter:2010hi,Borowka:2012yc,Borowka:2013cma},
here  we will only mention the main
aspects, and describe features not highlighted previously. \\
The {\it loop} part has been extended in version $2.1$ to be able to treat 
parametric integrals which are not in the canonical form as  
obtained directly from standard Feynman parametrisation. 
The integrals can have a different 
format, coming for example from variable transformations and/or  analytical 
integrations over some of the Feynman parameters. 
Hence, as these functions differ from the standard representation which the program would derive from 
the propagators in an automated way, they have to be defined in an input file by the user.
Contour deformation is available for these functions, as they are assumed to originate 
from a Feynman integral structure, where poles on the real axis are protected 
by the infinitesimal $i\delta$ prescription.

The setup in the directory {\it general} is designed to deal with even more general parametric
functions, where the integrand can consist of a product 
of arbitrary length of polynomial functions to some power. 
However, these functions should have only endpoint singularities
(i.e. dimensionally regulated singularities at the integration boundaries). 
Contour deformation is not available in this case because the 
correct sign of the imaginary part for the deformation into the complex plane 
cannot be inferred if the assumption of an underlying 
Feynman integral structure is dropped.

The procedure of iterated sector decomposition and subsequent numerical integration
is the same for all the different types of input functions, 
and is described in \cite{Carter:2010hi,Borowka:2012yc,Heinrich:2008si}.
\\
The current version 2.1.4 of the program can be downloaded from \\
{\tt http://projects.hepforge.org/secdec}.\\
Unpacking the tar archive 
will create a directory called {\tt SecDec-2.1.4}. 
Changing to this directory
and running {\it ./install} will compile the {\sc Cuba} library\,\cite{Hahn:2004fe} 
needed for the numerical integration. 
Prerequisites are Mathematica (version $\geq 6$), Perl (installed by default on 
most Unix/Linux systems) and a C++ compiler, respectively a Fortran compiler 
if the Fortran option is used.\\
More details about the usage, in particular about the option {\tt userdefined},
can be found in Refs.~\cite{Carter:2010hi,Borowka:2013cma}, 
and also in the documentation coming with the program.

\section{Applications}
\label{sec:results}
\subsection{Non-planar massive two-loop diagrams entering NNLO $t\bar{t}$ production}
\begin{figure}[h]
\subfigure[ggtt1]{\includegraphics[width=0.4\textwidth]{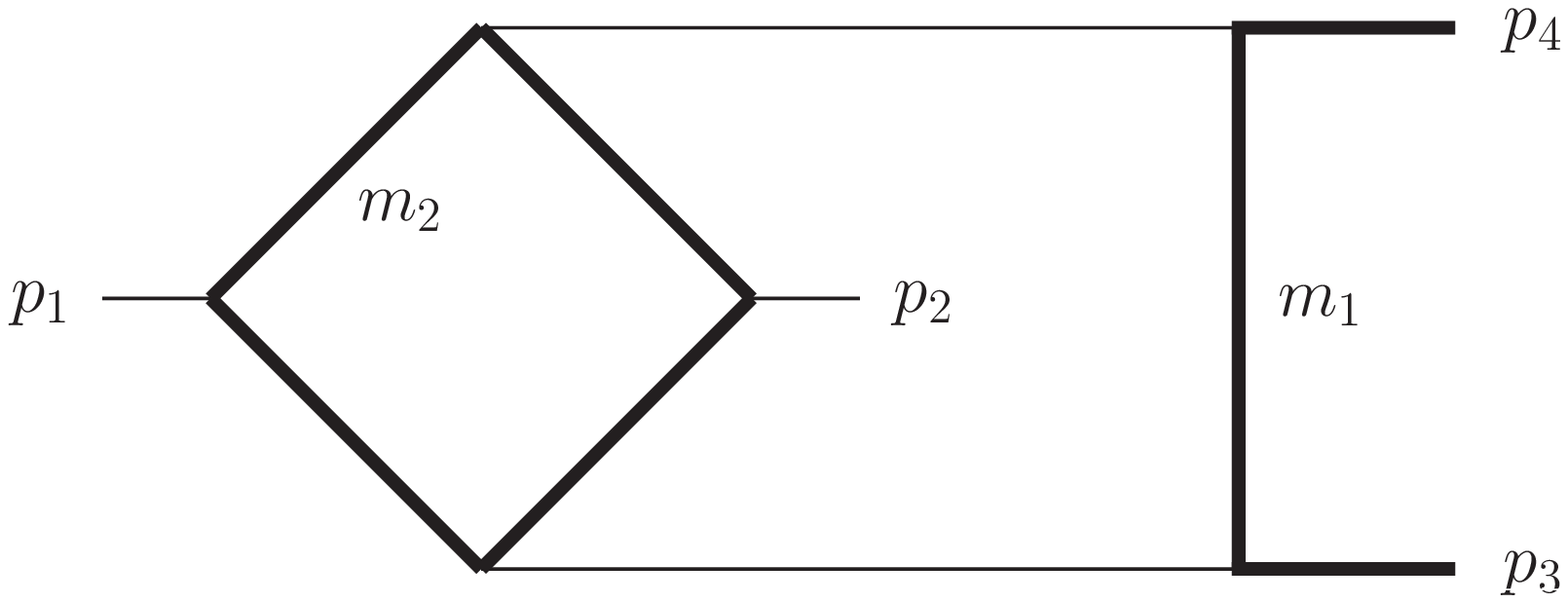}}
\hspace{2cm}
\subfigure[ggtt2]{\includegraphics[width=0.4\textwidth]{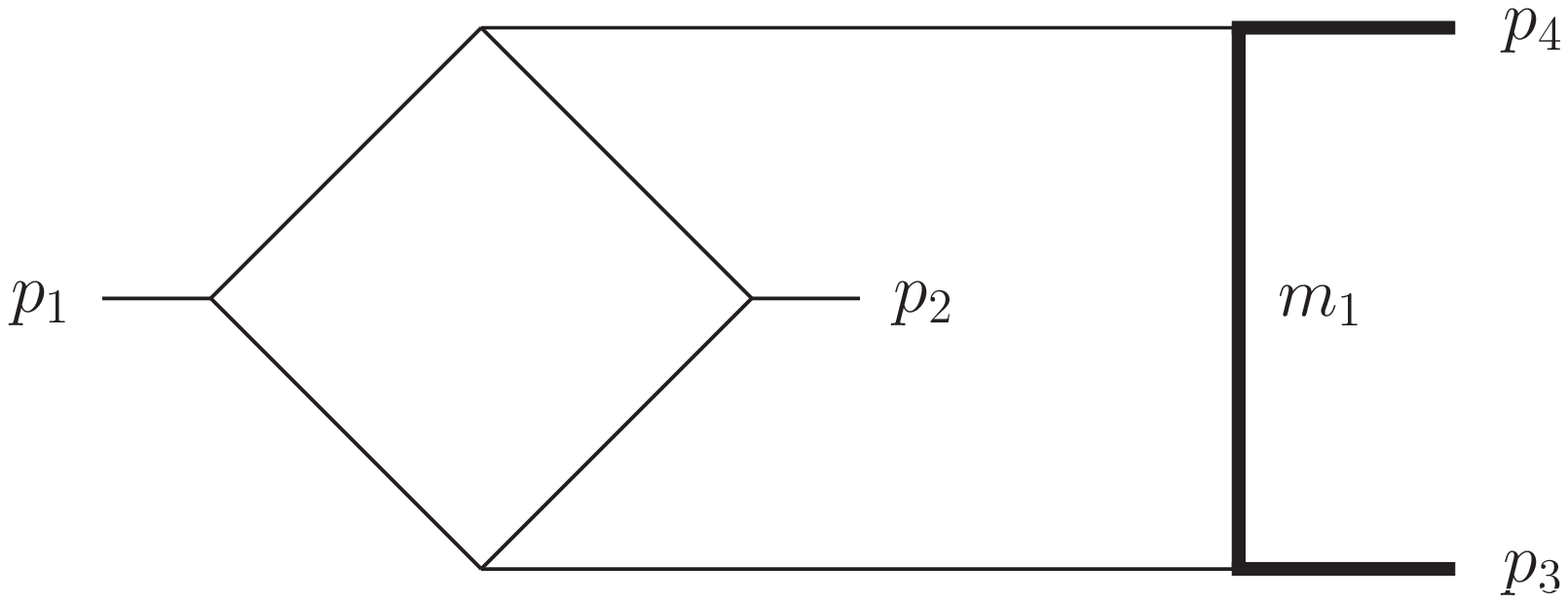} }
	\caption{Massive non-planar two-loop box diagrams entering the heavy (a) and 
	light (b) fermionic correction to the $gg\to t\bar{t}$ channel; the bold lines denote massive particles.} 
	\label{fig:ggtt2} 
\end{figure}

The most complicated master topologies occurring in the two-loop corrections 
to $t\bar{t}$ production in the $gg$ channel are the non-planar seven-propagator integrals shown in 
Fig.~\ref{fig:ggtt2}.

Analytic results for the integral 
containing a sub-diagram with a massive loop (called {\it ggtt1} here), are not available, 
while analytic results for 
the diagram corresponding to massless fermionic corrections 
in a sub-loop (called {\it ggtt2} here) have become available 
very recently\,\cite{vonManteuffel:2013uoa}.
However, the numerical evaluation of {\it ggtt1} with \secdec{} is 
much easier than the one of {\it ggtt2}, 
due to its less complicated infrared singularity structure. 
While the leading poles of {\it ggtt2} are of order $\mathcal{O}(\epsilon^{-4})$, 
and intermediate expressions during sector decomposition 
contain (spurious) poles where the degree of divergence is higher than logarithmic,
the integral {\it ggtt1} is finite and free from spurious singularities.
Therefore we can evaluate {\it ggtt1} with \secdec\,2.1 
using the fully automated setup.
In contrast,  for {\it ggtt2}
it turned out to be advantageous to make some analytical manipulations beforehand.
In particular, it was useful to perform one parameter integration analytically 
before feeding the integral into the decomposition and numerical integration algorithm. 
Further, we introduced special transformations to reduce the occurrence of spurious 
singularities, which are described in detail in \cite{Borowka:2013cma,Borowka:2013lda}.
These manipulations lead to functions which were not in the ``standard form" 
of Feynman parameterised loop integrals   anymore. 
This entailed the development of a setup to treat ``non-standard" 
parametric functions mentioned in Section \ref{sec:structure}, which has been made available for the user, 
as it can be beneficial in similar contexts.

\begin{figure}[htb]
\subfigure[finite part, including threshold]{
\includegraphics[width=6.5cm]{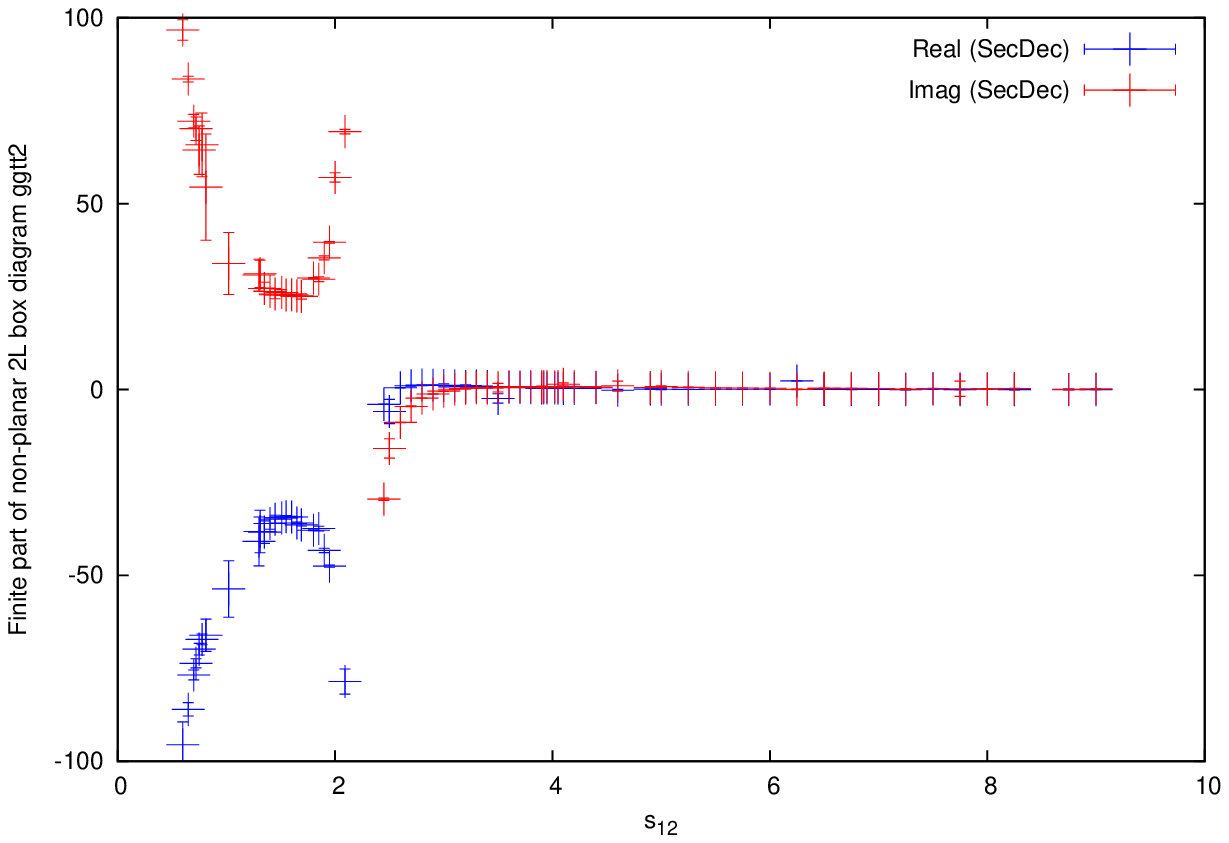} }
\hfill
\subfigure[finite part, beyond threshold]{
\includegraphics[width=6.5cm]{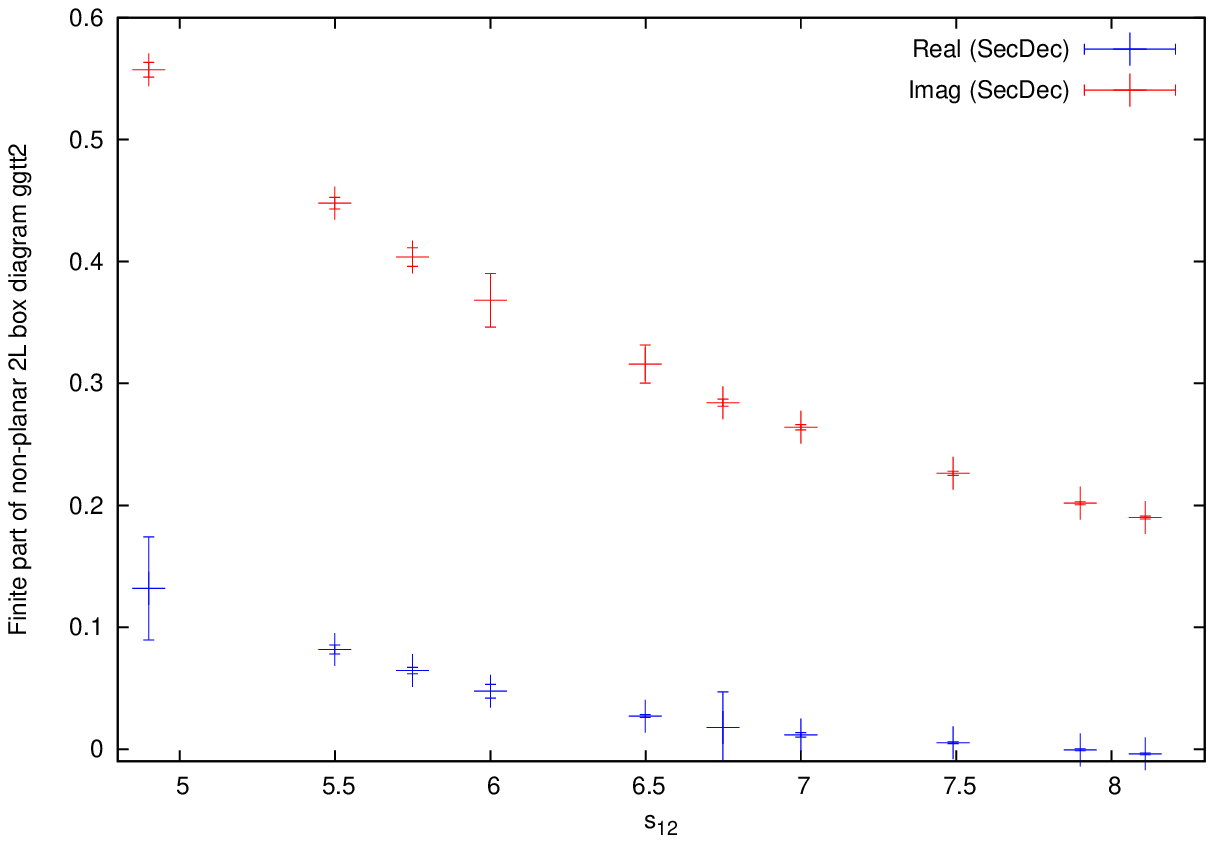} }
\caption{Results for the finite part of the scalar integral  {\it ggtt2}, 
(a) for a larger kinematic range, (b) zoom into a region further away from threshold.
 The vertical bars denote the numerical integration errors.}
\label{fig:ggtt2finite}
\end{figure}

Numerical results for the {\it ggtt2} diagram are shown in 
Fig.~\ref{fig:ggtt2finite}, 
where we used 
the numerical values $p_3^2=p_4^2=m^2=1, s_{23}=-1.25, s_{13}=2m^2-s_{12}-s_{23}$, 
and we extract an overall factor of $-16\,\,\Gamma(1+\eps)^2$. 
We only show results for the  finite part here, 
as it is the most complicated one and therefore more interesting than the pole coefficients.

\begin{figure}[htb]
\subfigure[scalar integral]{\includegraphics[width=7.cm]{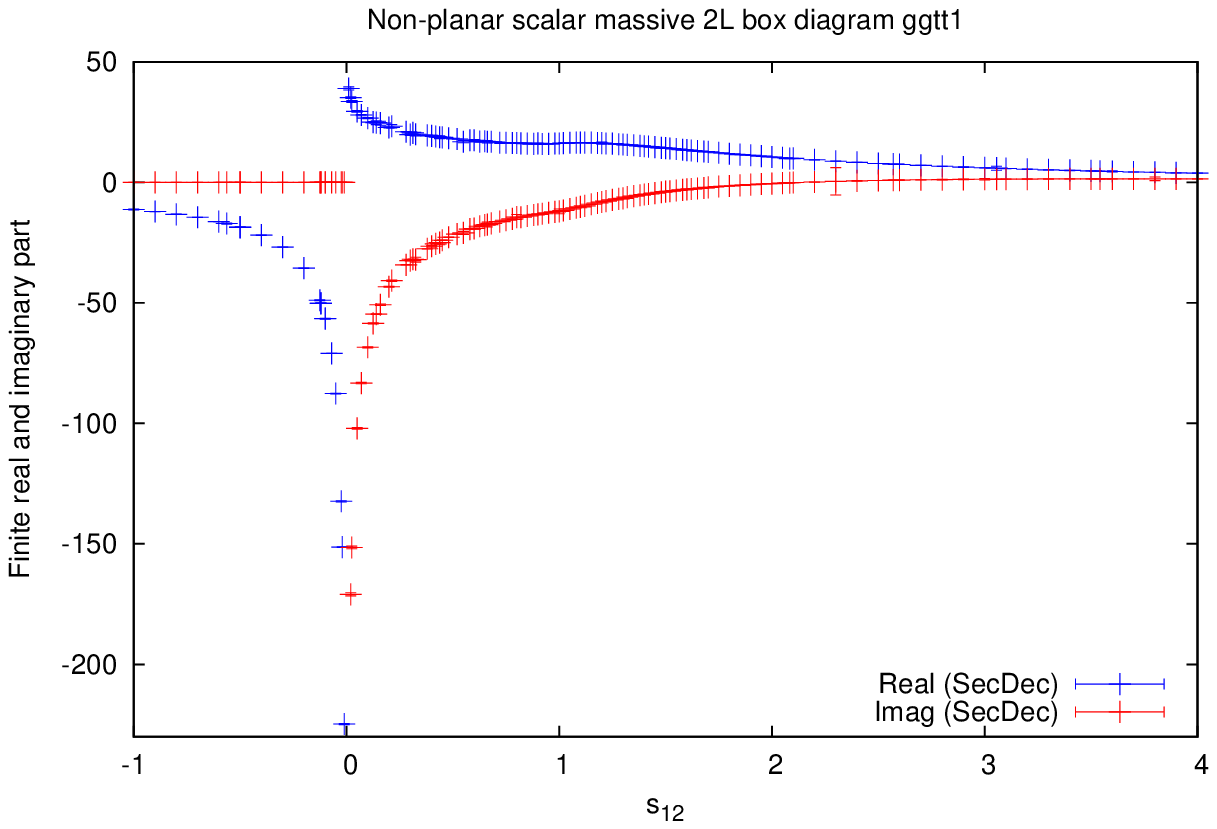} }\hfill
\subfigure[rank 2 tensor integral]{\includegraphics[width=7.cm]{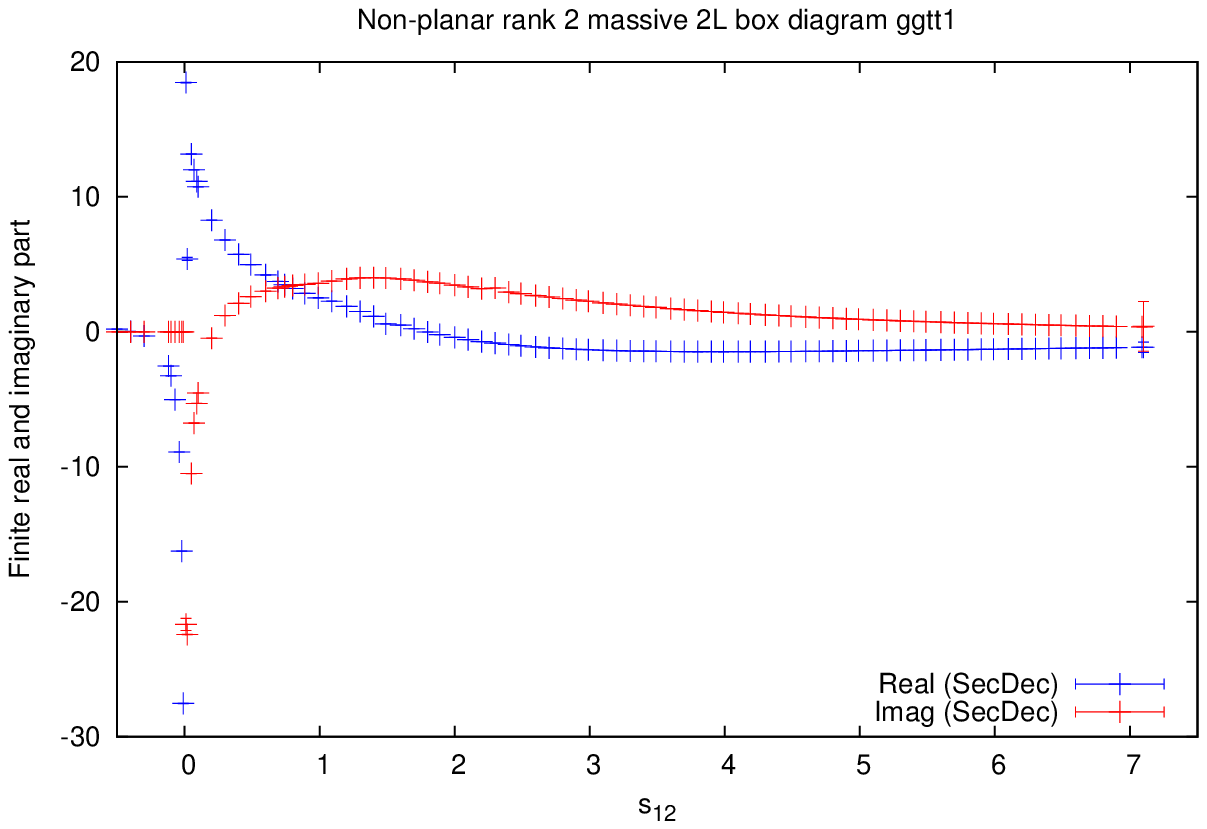} }
\caption{Results for the scalar integral  {\it ggtt1} 
and the corresponding rank two tensor integral  with $k_1\cdot k_2$ in the numerator. 
We vary $s_{12}$ and fix $s_{23}=-1.25, m_2=m_1, p_3^2=p_4^2=m_1^2=1$.}
\label{fig:ggtt1}
\end{figure}

Numerical results for the diagram {\it ggtt1} 
are shown in Fig.~\ref{fig:ggtt1} for both the scalar integral and an irreducible
rank two tensor integral. 

For the results shown in  Fig.~\ref{fig:ggtt1} we used 
the numerical values $m_1^2=m_2^2=m^2=1, s_{23}=-1.25, s_{13}=2\,m^2-s_{12}-s_{23}$.
We set $m_1^2=m_2^2$
because this is the only case occurring in the process $gg\to t\bar{t}$ at two loops
if the $b$-quarks are assumed to be massless. However, we also verified that the case 
$m_1^2\not=m_2^2$ can be evaluated without a significant increase in computing time compared to the 
equal mass case.


The timings for one kinematic point for the scalar integral in Fig.\,\ref{fig:ggtt1}(a) 
range from 11-60 secs for points far from threshold to $1.6\times 10^3$ 
seconds for a point very close to threshold, with an average of about 500 secs 
for points in the vicinity of the threshold.
A relative accuracy 
of $10^{-3}$ has been required for the numerical integration, 
while the absolute accuracy has been set to  $10^{-5}$. 
For the tensor integral, the timings are  better than in the scalar case, 
as the numerator function present in this case smoothes out the singularity structure. 
The timings were obtained on a single machine using Intel i7 processors and 8 cores.

\subsection{Corrections to the neutral CP even Higgs bosons in the MSSM}

The Higgs sector of the MSSM consists of two doublets $H_1$ and $H_2$, which can be written as 
\begin{align}
H_1 = \left( \begin{matrix} H_1^0 \\ H_1^- \end{matrix} \right) = 
\left( \begin{matrix} v_1 + \frac{1}{\sqrt{2}}(\phi_1^0 + \textrm{i} \chi_1^0) \\ -\phi_1^- \end{matrix} \right) \text{, } \hspace{10pt} 
H_2 = \left( \begin{matrix} H_2^+ \\ H_2^0 \end{matrix} \right) = 
\left( \begin{matrix}  \phi_2^+ \\ v_2 + \frac{1}{\sqrt{2}}(\phi_2^0 + \textrm{i} \chi_2^0) \end{matrix} \right) \text{.}
\end{align}
 The vacuum expectation values $v_1$ and $v_2$ define the angle $\tan\beta=v_2/v_1$.
At tree level, the mass matrix of the neutral CP-even Higgs bosons in the $\phi_1,\phi_2$ basis can be written as 
\begin{align}
\nonumber M_{\text{Higgs}}^{2,\text{tree}}=\left( \begin{matrix} M_A^2\text{sin}^2\, \beta + M_Z^2\text{cos}^2\, \beta & -(M_A^2 + M_Z^2)\,\text{sin}\, \beta \text{cos}\, \beta \\ -(M_A^2 + M_Z^2)\,\text{sin}\, \beta \text{cos}\, \beta & M_A^2\text{cos}^2\, \beta + M_Z^2\text{sin}^2\, \beta \end{matrix} \right)\;,
\end{align}
where $M_A$ is the mass of the CP-odd neutral Higgs boson $A$. 
The rotation to the basis formed by the Higgs bosons $H^0,h^0$ is given by 
\begin{align}
 \left( \begin{matrix} H^0 \\ h^0 \end{matrix} \right) =  \left( \begin{matrix} \textrm{cos} \,\alpha  & \textrm{sin} \,\alpha \\ 
-\textrm{sin} \,\alpha & \textrm{cos} \, \alpha \end{matrix} \right) \left( \begin{matrix} \phi_1^0 \\ \phi_2^0 \end{matrix} \right)\;,
\end{align}
where 
\begin{align}
 \textrm{tan} (2 \alpha) =  \textrm{tan} (2 \beta) \frac{M_A^2 + M_Z^2}{M_A^2-M_Z^2} \text{ ,} \hspace{20pt} -\frac{\pi}{2} < \alpha < 0 \text{ .}
\end{align}
The need for higher order corrections to the MSSM Higgs boson masses is obvious from the fact that at tree level, 
the mass of light CP-even Higgs boson is bound from above through the relation $m_h \leq \text{min}(M_Z,M_A) \,|\text{cos}(2 \beta)|$.\\
The status of currently available two-loop self-energy corrections to the Higgs boson masses in the MSSM with real parameters (rMSSM) is the following. 
The corrections $\mathcal{O}(\alpha_s \alpha_t)$\,\cite{Hempfling:1993qq,Heinemeyer:1998jw,Heinemeyer:1998kz,Heinemeyer:1998np,Espinosa:1999zm,Degrassi:2001yf}, $\mathcal{O}(\alpha_t^2)$\,\cite{Hempfling:1993qq,Espinosa:2000df,Brignole:2001jy}, 
$\mathcal{O}(\alpha_s \alpha_b)$\,\cite{Brignole:2002bz}, $\mathcal{O}(\alpha_t \alpha_b)$\,\cite{Dedes:2003km}, 
$\mathcal{O}(\alpha_b^2)$\,\cite{Dedes:2003km} are known in the gaugeless limit (i.e. vanishing gauge couplings to the 
SM vector bosons), and 
using the approximation $p^2=0$ for the external momentum of the self-energies. 
The strong coupling constant is denoted by $\alpha_s$ as usual, while $\alpha_{\{t,b\}}$ denote the Yukawa couplings 
of the top (bottom) quarks, where 
$\alpha_{\{t,b\}}= y^2_{\{t,b\}}/(4\pi)$.
At three loops, the $\mathcal{O}(\alpha_s^2 \alpha_t)$ corrections are known~\cite{Martin:2007pg,Harlander:2008ju,Kant:2010tf} 
in the gaugeless limit and in the $p^2=0$ approximation.
The largest uncertainties on the presently known results stem from the contributions of the momentum dependent self-energy corrections at two loop order. 
The latter have been calculated in \cite{Martin:2002wn,Martin:2004kr} based on an effective potential approach within the $\overline{DR}$ scheme. 
However, in order to incorporate the corrections into the public program {\tt FeynHiggs}\,\cite{Hahn:2009zz,Hahn:2010te}, 
it is more convenient to use on-shell renormalisation for all the masses.
To this aim, we computed the self-energies $\Sigma_{\phi_i\phi_j}(p^2)$ at two loops at  $\mathcal{O}(\alpha_s \alpha_t)$, 
keeping the full momentum dependence. We use the $\overline{DR}$ scheme for the field renormalisation,
and on-shell renormalisation for all other renormalisation constants, as explained e.g. in \cite{Heinemeyer:1998np}.

\begin{figure}[htb]
\subfigure[T234]{\includegraphics[width=0.23\textwidth]{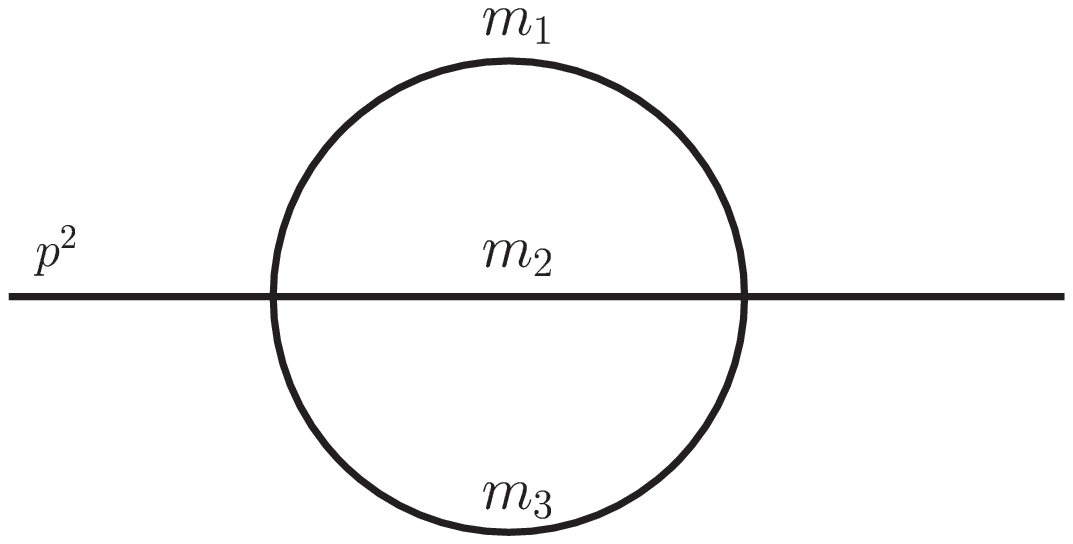}}\hspace{1pt}
\subfigure[T1234]{\includegraphics[width=0.23\textwidth]{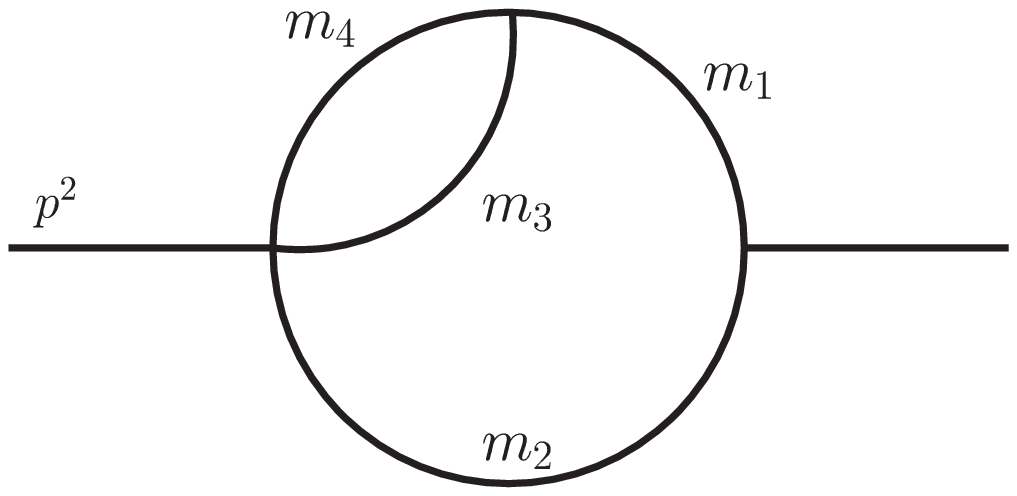}}\hspace{1pt}
\subfigure[T11234]{\includegraphics[width=0.23\textwidth]{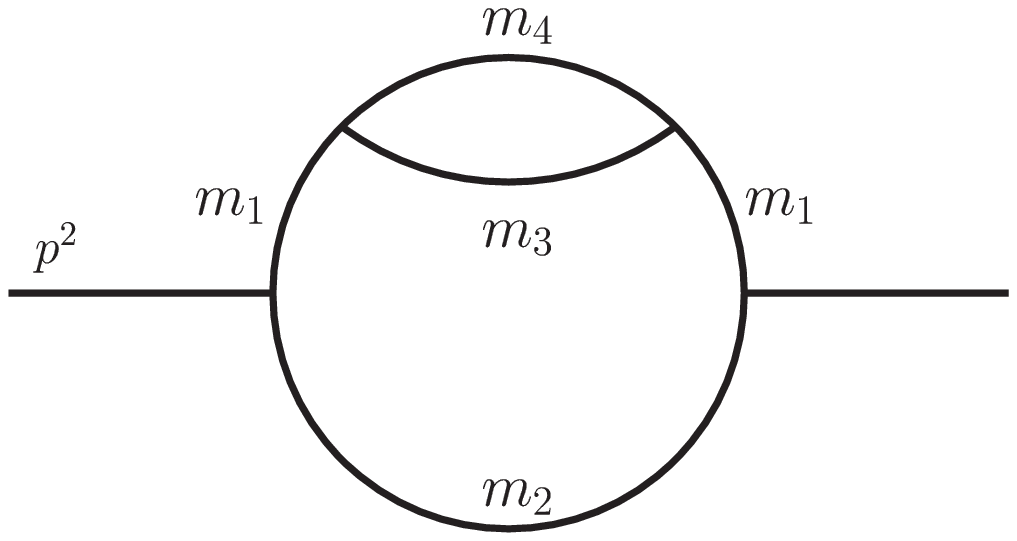}}\hspace{1pt}
\subfigure[T12345]{\includegraphics[width=0.23\textwidth]{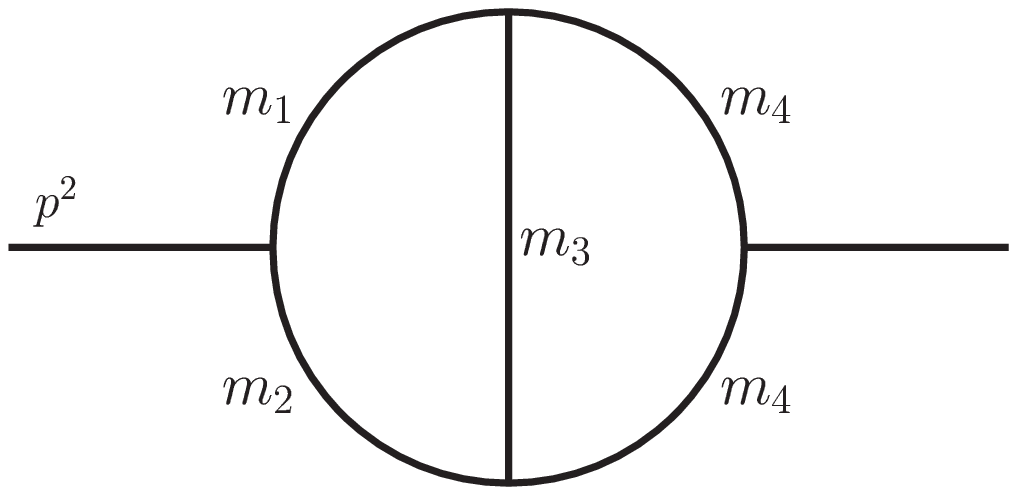}}
\caption{Two-loop two-point topologies treated with \secdec{} with up to four different mass scales}
\label{fig:Tnum}
\end{figure}

\begin{figure}[htb]
\subfigure[]{\includegraphics[height=5.5cm]{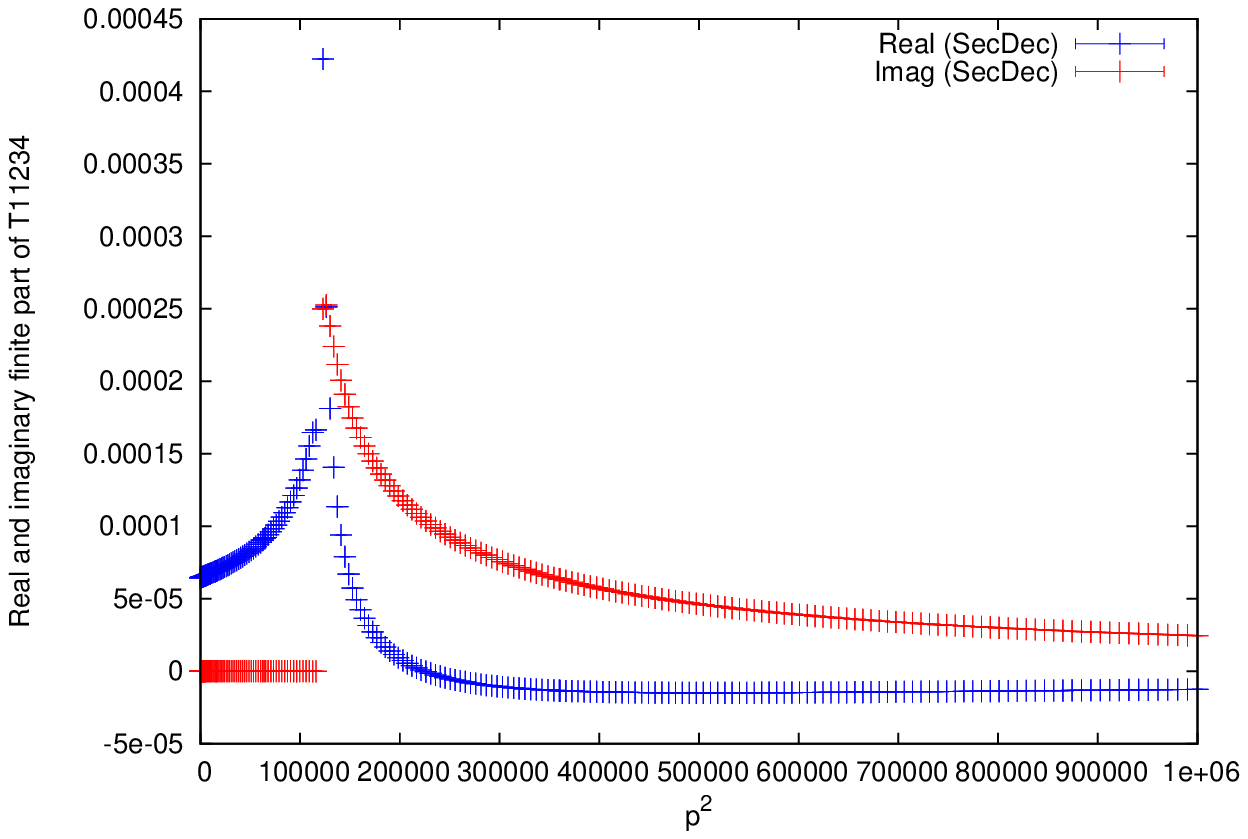} \label{fig:T11234} }\hfill
\subfigure[]{\includegraphics[height=5.5cm]{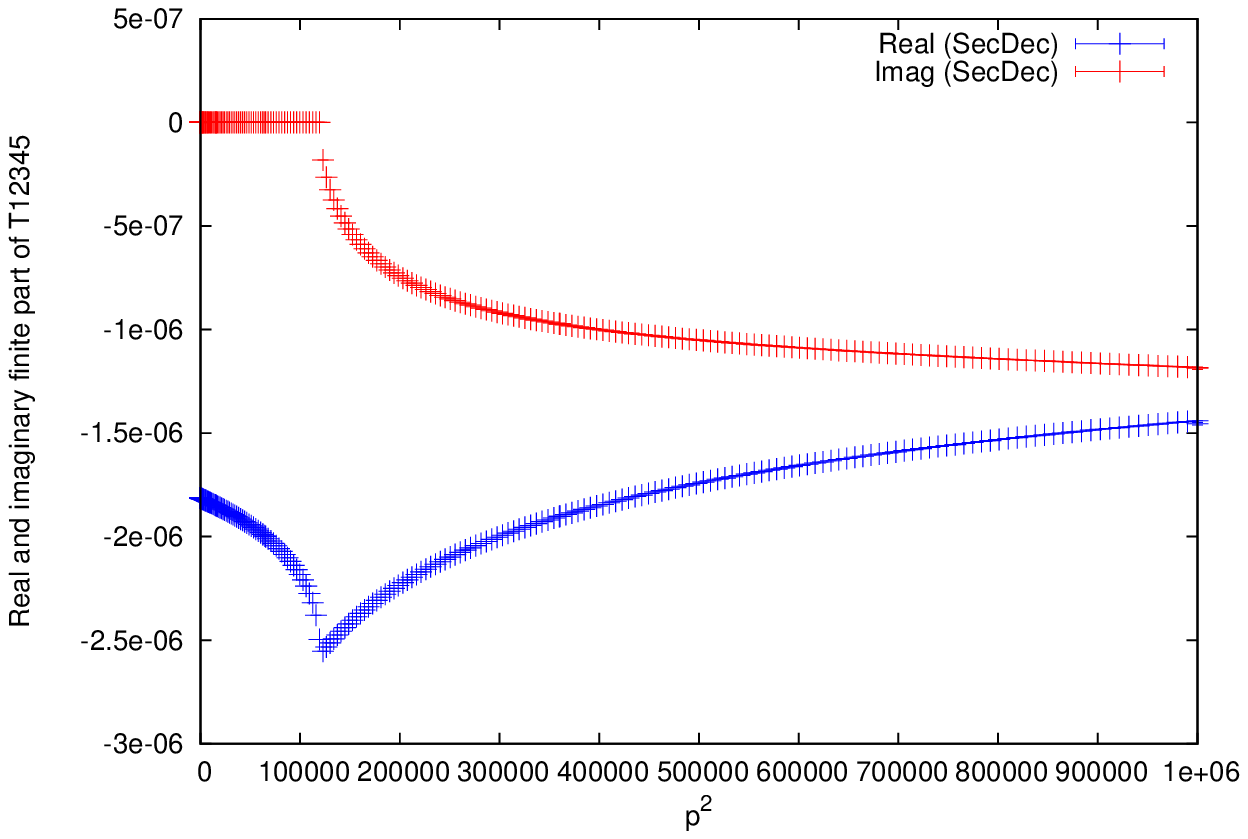} \label{fig:T12345} }
\caption{Numerical \secdec{} results for the integrals of type (c) and (d) as defined in Fig. 5.}
\label{fig:Tints}
\end{figure}

The reduction of the amplitude to scalar master integrals is performed with the help of the 
package {\tt TwoCalc}\,\cite{Weiglein:1993hd} for the two-loop diagrams, while the one-loop counter term diagrams are reduced using {\tt FormCalc} \cite{Agrawal:2012cv}.
The two-loop topologies for which a full analytical result is not available, shown in Fig.\ref{fig:Tnum}, 
are calculated numerically with \secdec.
In total, these amount to 32 different mass configurations, where the two-loop diagrams can have up to four different masses. 
We tested the numerical integration with \secdec{} for integrals where the kinematic values  differed by up to 14 orders of magnitude. 
The timings range between 0.01 secs and 100 secs with an achieved relative accuracy of $10^{-5}$ to $10^{-11}$.
Two representative results are shown in Fig. \ref{fig:Tints}. The configuration in Fig. \ref{fig:T11234} has three different mass scales where we set $m_1 = m_2 =173.2$ GeV, $m_3 =826.8$ GeV and $m_4  =1.5$ TeV. In Fig. \ref{fig:T12345}, the numerical values $m_1 =1173.2$ GeV, $m_2 =826.8$ GeV, $m_3 =1.5$ TeV and $m_4 =173.2$ GeV are chosen.\\
More detailed results will be given elsewhere\,\cite{toappear}.

\section{Conclusions}

We have presented new features and applications of the program \secdec, 
which can be used to 
calculate multi-loop integrals numerically in an automated way. 
Applications to non-planar two-loop master integrals occurring in 
$t\bar{t}$ production are shown, as well as the calculation of 
momentum dependent two-loop corrections to the masses of the 
neutral CP-even Higgs bosons in the MSSM.
The program \secdec{} is publicly available at {\tt http://projects.hepforge.org/secdec}.

\subsection*{Acknowledgments}
We would like to thank Andreas von Manteuffel for comparisons with analytic results. We also 
thank Wolfgang Hollik for the productive collaboration on the calculation of the corrections to 
the MSSM Higgs masses and Sven Heinemeyer for interesting discussions. 
SB also wants to thank the organizers of RADCOR 2013 for the nice conference.

\bibliography{refs_secdec}

\end{document}